\documentclass[prd,nofootinbib,a4paper,showkeys,preprintnumbers,twocolumn]{revtex4}

\usepackage{amsmath}
\usepackage{amssymb}
\usepackage{amsfonts}
\usepackage{latexsym}
\usepackage{mathrsfs}

\begin{document}

\title{A remark on the Koide relation for quarks}

\author{A. Kartavtsev$^{a}$}
\email[\,]{alexander.kartavtsev@mpi-hd.mpg.de}

\affiliation{Max-Planck-Institut f\"ur Kernphysik, Saupfercheckweg 1, 69117 Heidelberg, Germany}

\begin{abstract}
The charged lepton masses obey to high precision the so-called Koide relation. We propose  a generalization
of this relation to quarks. It  includes up and down quarks of the three generations and is numerically reasonably
close to the Koide limit. 
\end{abstract}

\maketitle

The pole masses of three charged leptons have been measured to an unprecedented precision \cite{PDG}:
\begin{subequations}
\label{LeptonMassesLowScale}
\begin{align}
m_e&=0.510998910 \pm 0.000000013\,\, {\rm MeV}\,,\\
m_\mu&=105.658367 \pm 0.000004\,\, {\rm MeV}\,,\\
m_\tau&=1776.82^{+0.16}_{-0.26}\,\, {\rm MeV}\,.
\end{align}
\end{subequations}
To a surprising degree of accuracy they obey the empirical Koide mass relation \cite{Koide:1982si,koide_new_1983}:
\begin{align}
\label{KoideLeptons}
K_\ell\equiv \frac{m_e+m_ \mu+m_\tau}{(\sqrt{m_e}+\sqrt{m_\mu}+\sqrt{m_\tau})^2}=\frac23\,.
\end{align}
Defining a  deviation from the Koide limit, $\Delta K_\ell\equiv (3 K_\ell/2-1)$, and substituting the experimental 
charged lepton masses into \eqref{KoideLeptons} we  find $-2\cdot 10^{-5} < \Delta K_\ell < 9\cdot 10^{-7}$. It is 
also astonishing that the electron mass,  being orders of magnitude smaller than the muon and tau-lepton masses, 
still plays an important role in this relation. Neglecting $m_e$ in \eqref{KoideLeptons} we would find $\Delta K_\ell \sim 2 \cdot 10^{-2}$.

There have been quite a few attempts to derive \eqref{KoideLeptons} from flavor symmetries or grand-unified 
extensions of the Standard Model \cite{Koide:1989jq,koide_analytical_1996,koide_u3-family_1995}. Inspired 
by the idea of grand unification, one may wonder whether a similar mass relations also exist for quarks. Using 
current values of $d$, $s$ and $b$--quark masses \cite{PDG}:
\begin{subequations}
\label{DownQuarkMasses}
\begin{align}
m_d&=4.1 - 5.7\,\, {\rm MeV}\,,\\
m_s&=100^{+30}_{-20}\,\, {\rm MeV}\,,\\
m_b&=4.19^{+0.18}_{-0.06}\,\, {\rm GeV}\,,
\end{align}
\end{subequations}
and defining, analogously to \eqref{KoideLeptons}, 
\begin{align}
\label{KoideDown}
K_{d}\equiv \frac{m_d+m_s+m_b}{(\sqrt{m_d}+\sqrt{m_s}+\sqrt{m_b})^2}\,,
\end{align}
we find  $K_{d}\sim 0.72$, i.e. a considerable deviation from the Koide limit with $\Delta K_d \sim 8 \cdot 10^{-2}$.
For up-type quarks the mass spectrum has a stronger hierarchy than for down-type quarks \cite{PDG}:
\begin{subequations}
\label{UpQuarkMasses}
\begin{align}
m_u&=1.7 - 3.1\,\, {\rm MeV}\,,\\
m_c&=1.29^{+0.05}_{-0.11}\,\, {\rm GeV}\,,\\
m_t&=172.9 \pm 0.6 \pm 0.9 \,\, {\rm GeV}\,,
\end{align}
\end{subequations}
and therefore the analogue of \eqref{KoideLeptons} for up-type quarks,  
\begin{align}
\label{KoideUp}
K_{u}\equiv \frac{m_u+m_c+m_t}{(\sqrt{m_u}+\sqrt{m_c}+\sqrt{m_t})^2}\,,
\end{align}
deviates from the Koide limit even stronger, $K_{u}\sim 0.85$, which corresponds to $\Delta K_u  \sim 27 \cdot 10^{-2}$.
One could argue, that due to the non-perturbative nature of quantum chromodynamics at low energies, the pole masses 
of light quarks are not well defined and one should make use of the running quark masses at the scale $M_Z = 91.2$ GeV \cite{rodejohann_extended_2011}. The latter are given by \cite{xing_updated_2007}:
\begin{subequations}
\label{QuarkMassesAtMZ}
\begin{align}
m_u&=1.27^{+0.50}_{-0.42}\,\, {\rm MeV}\,,\\
m_d&=2.90^{+1.24}_{-1.19}\,\, {\rm MeV}\,,\\
m_s&=55^{+16}_{-15}\,\, {\rm MeV}\,,\\
m_b&=0.619\pm 0.084\,\, {\rm GeV}\,,\\
m_c&=2.89\pm 0.09\,\, {\rm GeV}\,,\\
m_t&=171.7\pm 3.0\,\, {\rm GeV}\,.
\end{align}
\end{subequations}
Substituting \eqref{QuarkMassesAtMZ} into \eqref{KoideDown} and \eqref{KoideUp} we find 
$K_{d}\sim 0.74$ and $K_{u}\sim 0.89$ respectively. This corresponds to $\Delta K_d  \sim 11 \cdot 10^{-2}$
and $\Delta K_u  \sim 33 \cdot 10^{-2}$. That is, the running effects induce an even stronger deviation
from the Koide limit. The same also applies to leptons.  The running charged lepton masses at the scale $M_Z = 91.2$ 
GeV are given by \cite{xing_updated_2007}:
\begin{subequations}
\begin{align}
m_e&=0.486570161 \pm 0.000000042\,\, {\rm MeV}\,,\\
m_\mu&=102.7181359 \pm 0.0000092\,\, {\rm MeV}\,,\\
m_\tau&=1746.24^{+0.20}_{-0.19}\,\, {\rm MeV}\,.
\end{align}
\end{subequations}
Substituting these values into the Koide relation for leptons \eqref{KoideLeptons}  we  find 
$1.87 \cdot 10^{-3} < \Delta K_\ell  < 1.91\cdot 10^{-3}$. This implies in particular, that the 
running masses at high scales are not compatible with the exact Koide limit, $K_\ell=2/3$. This result  is somewhat 
counterintuitive. If there is an underlying high-scale symmetry which ensures \eqref{KoideLeptons} 
for leptons, one could expect that the deviation from the Koide limit would be smaller at 
high scales. 

It was noted  in \cite{rodejohann_extended_2011} that one could  divide quarks into light and heavy ones, 
instead of up- and down-like quarks, that is, according to their mass instead of the  isospin.
The corresponding Koide parameters are defined by: 
\begin{subequations}
\label{KoideLightHeavy}
\begin{align}
K_{light}&\equiv \frac{m_u+m_d+m_s}{(\sqrt{m_u}+\sqrt{m_d}+\sqrt{m_s})^2}\,,\\
K_{heavy}&\equiv \frac{m_c+m_b+m_t}{(\sqrt{m_c}+\sqrt{m_b}+\sqrt{m_t})^2}\,.\
\end{align}
\end{subequations}
Substituting  \eqref{DownQuarkMasses} and \eqref{UpQuarkMasses} into \eqref{KoideLightHeavy} we find 
a rather good agreement with the Koide limit for the heavy quarks,  $-5 \cdot 10^{-3} < \Delta K_{heavy} < 1 \cdot 10^{-2}$,
and a deviation from it for the light ones,  $-20 \cdot 10^{-2} < \Delta K_{light} < - 6 \cdot 10^{-2}$ . If we take  
the running masses at the scale $M_Z = 91.2$ GeV instead, we will obtain $5 \cdot 10^{-2} < \Delta K_{heavy} < 9 \cdot 10^{-2}$  and  
$-20 \cdot 10^{-2} < \Delta K_{light} < - 2 \cdot 10^{-2}$ respectively.

Although  dividing quarks into the light and the heavy ones numerically gives a rather good agreement with the 
Koide relation, the fact that $s$ and $c$ quarks -- components of the same  $SU_L(2)$ doublet -- 
enter the expressions for $K_{light}$ and $K_{heavy}$ separately is counterintuitive. A somewhat more natural 
generalization of the Koide relation \eqref{KoideLeptons} to quarks is an expression which sums over 
up \textit{and} down components of the three generations:
\begin{align}
\label{KoideAll}
K_{q}\equiv \frac{\sum m_q}{\left(\sum\sqrt{m_q}\right)^2}\,.
\end{align}
Since the active neutrino masses are very small, a similar generalization of the Koide relations for 
leptons would not affect the range for $\Delta K_\ell$  indicated above. 
Substituting  \eqref{DownQuarkMasses} and \eqref{UpQuarkMasses} into \eqref{KoideAll} we find for 
the deviation from the Koide limit $-5 \cdot 10^{-2} < \Delta K_q < - 4 \cdot 10^{-2}$. If we take  the 
running quark masses at the scale $M_Z = 91.2$ GeV instead, we will obtain 
$2 \cdot 10^{-2} < \Delta K_q < 5 \cdot 10^{-2}$. It is interesting that due to the running effects 
$\Delta K_q$ crosses zero. That is, at some scale 2 GeV $< \mu < M_Z$  the quark masses 
satisfy (within the uncertainties) the Koide relation $K_q=2/3$. 

It was noted in \cite{foot_note_1994}  that the Koide relation has a geometric interpretation. Let 
$(\sqrt{m_e},\sqrt{m_\mu},\sqrt{m_\tau})$ define a vector in a three-dimensional vector space and 
$\theta_\ell$ be the angle between this vector and the vector $(1,1,1)$,
\begin{align}
\label{theta_ell}
\theta_\ell\equiv \arccos\frac{(\sqrt{m_e},\sqrt{m_\mu},\sqrt{m_\tau})\cdot(1,1,1)}{|(\sqrt{m_e},\sqrt{m_\mu},\sqrt{m_\tau})||(1,1,1)|}\,.
\end{align}
Then the Koide relation \eqref{KoideLeptons} is equivalent to the statement $\theta_\ell=\pi/4$. Any deviation 
of $K$ from the Koide limit, $K=2/3$, implies also a deviation from $\theta=\pi/4$. Thus, $\theta_d$ and $\theta_d$
substantially deviate from $\pi/4$, whereas $\theta_{light}$ and $\theta_{heavy}$ are close to this limit. It is 
interesting that if we define an analogue of \eqref{theta_ell} for quarks in a six-dimensional vector space,
\begin{align}
\theta_q\equiv \arccos\frac{(\sqrt{m_d},\sqrt{m_u},\sqrt{m_s}, \ldots)\cdot(1,1,1,\ldots)}{|(\sqrt{m_d},\sqrt{m_u},\sqrt{m_s},\ldots)||(1,1,1,\ldots)|}\,,
\end{align}
then we will find that this angle is  again a rational multiple of $\pi$, $\theta_q\approx\pi/3$. 

To summarize,  generalizations of the original Koide relation to quarks which include only the  up-type or only the 
down-type quarks show a considerable deviation from the Koide limit $K=2/3$. In this letter we have proposed 
a generalization  which includes up \textit{and} down quarks of the three generations. Interestingly enough, 
although masses of the light ($u$, $d$ and $s$) quarks are much smaller than masses of the heavy ones 
($c$, $b$ and $t$), their inclusion into the sum makes it substantially closer to the Koide limit. This resembles 
the role played by the electron mass in the Koide relation for charged leptons. Furthermore, similarly to the 
Koide relation for leptons, the proposed relation for quarks also allows for a geometric interpretation: the angle 
between vectors $(\sqrt{m_d},\sqrt{m_u},\sqrt{m_s}, \ldots)$ and $(1,1,1,\ldots)$ in a six-dimensional vector space is 
very close to $\pi/3$.

\subsection*{Acknowledgements}
\noindent
This work was supported by Deutsche Forschungsgemeinschaft under Grant KA-3274/1-1. 
The author would like to thank A. Hohenegger for careful reading of the manuscript. 




\begin{thebibliography}{9}
\expandafter\ifx\csname natexlab\endcsname\relax\def\natexlab#1{#1}\fi
\expandafter\ifx\csname bibnamefont\endcsname\relax
  \def\bibnamefont#1{#1}\fi
\expandafter\ifx\csname bibfnamefont\endcsname\relax
  \def\bibfnamefont#1{#1}\fi
\expandafter\ifx\csname citenamefont\endcsname\relax
  \def\citenamefont#1{#1}\fi
\expandafter\ifx\csname url\endcsname\relax
  \def\url#1{\texttt{#1}}\fi
\expandafter\ifx\csname urlprefix\endcsname\relax\def\urlprefix{URL }\fi
\providecommand{\bibinfo}[2]{#2}
\providecommand{\eprint}[2][]{\url{#2}}

\bibitem[{\citenamefont{Nakamura and Group}(2010)}]{PDG}
\bibinfo{author}{\bibfnamefont{K.}~\bibnamefont{Nakamura}} \bibnamefont{and}
  \bibinfo{author}{\bibfnamefont{P.~D.} \bibnamefont{Group}},
  \bibinfo{journal}{Journal of Physics G} \textbf{\bibinfo{volume}{37}},
  \bibinfo{pages}{075021} (\bibinfo{year}{2010}).

\bibitem[{\citenamefont{Koide}(1982)}]{Koide:1982si}
\bibinfo{author}{\bibfnamefont{Y.}~\bibnamefont{Koide}},
  \bibinfo{journal}{Lett.Nuovo Cim.} \textbf{\bibinfo{volume}{34}},
  \bibinfo{pages}{201} (\bibinfo{year}{1982}).

\bibitem[{\citenamefont{Koide}(1983)}]{koide_new_1983}
\bibinfo{author}{\bibfnamefont{Y.}~\bibnamefont{Koide}},
  \bibinfo{journal}{Phys.Rev. D} \textbf{\bibinfo{volume}{28}},
  \bibinfo{pages}{252} (\bibinfo{year}{1983}).

\bibitem[{\citenamefont{Koide}(1990)}]{Koide:1989jq}
\bibinfo{author}{\bibfnamefont{Y.}~\bibnamefont{Koide}},
  \bibinfo{journal}{Mod.Phys.Lett.} \textbf{\bibinfo{volume}{A5}},
  \bibinfo{pages}{2319} (\bibinfo{year}{1990}).

\bibitem[{\citenamefont{Koide and Fusaoka}(1996)}]{koide_analytical_1996}
\bibinfo{author}{\bibfnamefont{Y.}~\bibnamefont{Koide}} \bibnamefont{and}
  \bibinfo{author}{\bibfnamefont{H.}~\bibnamefont{Fusaoka}},
  \bibinfo{journal}{{hep-ph/9612322}}  (\bibinfo{year}{1996}),
  \bibinfo{note}{{Prog.Theor.Phys.} {\bf 97} (1997) 459-478}.

\bibitem[{\citenamefont{Koide and Tanimoto}(1995)}]{koide_u3-family_1995}
\bibinfo{author}{\bibfnamefont{Y.}~\bibnamefont{Koide}} \bibnamefont{and}
  \bibinfo{author}{\bibfnamefont{M.}~\bibnamefont{Tanimoto}},
  \bibinfo{journal}{{arXiv:9505333}}  (\bibinfo{year}{1995}),
  \bibinfo{note}{{Z.Phys.} C{\bf 72} (1996) 333-344}.

\bibitem[{\citenamefont{Rodejohann and Zhang}(2011)}]{rodejohann_extended_2011}
\bibinfo{author}{\bibfnamefont{W.}~\bibnamefont{Rodejohann}} \bibnamefont{and}
  \bibinfo{author}{\bibfnamefont{H.}~\bibnamefont{Zhang}},
  \bibinfo{journal}{Phys.Lett.} \textbf{\bibinfo{volume}{B {\bf 698}}},
  \bibinfo{pages}{152} (\bibinfo{year}{2011}), \eprint{arXiv:1101.5525}.

\bibitem[{\citenamefont{Xing et~al.}(2007)\citenamefont{Xing, Zhang, and
  Zhou}}]{xing_updated_2007}
\bibinfo{author}{\bibfnamefont{Z.-z.} \bibnamefont{Xing}},
  \bibinfo{author}{\bibfnamefont{H.}~\bibnamefont{Zhang}}, \bibnamefont{and}
  \bibinfo{author}{\bibfnamefont{S.}~\bibnamefont{Zhou}},
  \bibinfo{journal}{{arXiv:0712.1419}}  (\bibinfo{year}{2007}),
  \bibinfo{note}{{Phys.Rev. D {\bf 77} 113016}, 2008}.

\bibitem[{\citenamefont{Foot}(1994)}]{foot_note_1994}
\bibinfo{author}{\bibfnamefont{R.}~\bibnamefont{Foot}},
  \bibinfo{journal}{{arXiv:hep-ph/9402242}}  (\bibinfo{year}{1994}).

\end{thebibliography}
\end{document}